# ENTROPIC SELF-ASSEMBLY OF FREELY ROTATING POLYHEDRAL PARTICLES CONFINED TO A FLAT INTERFACE

V. Thapar, T. Hanrath, and F. A. Escobedo*
School of Chemical and Biomolecular Engineering, Cornell University
Ithaca New York 14853

AUTHOR EMAIL ADDRESS (*fe13@cornell.edu)

## Abstract

The self-assembly of hard polyhedral particles confined to a flat interface is studied using Monte Carlo simulations. The particles are pinned to the interface by restricting their movement in the direction perpendicular to it while allowing their free rotations. The six different polyhedral shapes studied in this work are selected from a family of truncated cubes defined by a truncation parameter, $s$, which varies from cubes ($s = 0$) via cuboctahedra ($s = 0.5$) to octahedra ($s = 1$). Our results suggest that shapes with small values of $s$ show square-like behavior whereas shapes with large values of $s$ tend to show more disc-like behavior. At an intermediate value of $s = 0.4$, the phase behavior of the system shows both square-like and disc-like features. The results are also compared with the phase behavior of 3D bulk polyhedra and of 2D rounded hard squares. Both comparisons reveal key similarities in the number and sequence of mesophases and solid phases observed. These insights on 2D entropic self-assembly of polyhedral particles is a first step toward understanding the self-assembly of particles at fluid-fluid interfaces, which is driven by a complex interplay of entropic and enthalpic forces.

## 1. Introduction

At present, a major focus in material science is the use of polyhedral colloidal nanoparticles as versatile building blocks to design and fabricate novel materials with targeted emergent properties. Remarkable strides have been made in experimental techniques[1-6] to synthesize these building blocks with different sizes, shapes, compositions and patterns. Understanding the relationship between particle shape and structural order is crucial for materials design because of the dependence of physical properties on the structure. This has fueled many theoretical[7,8] and simulation studies[9-21] of polyhedral particles to map out their phase behavior. These systems often undergo order-disorder phase transitions involving changes in both translational and orientational degrees of freedom and can lead to novel mesophases. A mesophase is a partially ordered phase whose properties are intermediate between those of disordered liquids and ordered crystals, such as liquid crystals, rotator plastic crystals, and quasicrystals.

Some of the techniques to assemble polyhedral nanoparticles involve the use of fluid interfaces as templates. The idea of using fluid interfaces was first explored for micron-sized particles by Pickering[22] and Ramsden[23] when they investigate paraffin water emulsions with solid particles of various oxides, salts, and clays. They found that these colloids generate a resistant film at the paraffin-water interface inhibiting the coalescence of the emulsion drops. These so-called Pickering emulsions are formed by the self-assembly of colloidal particles at fluid-fluid interfaces[24]. During these past years, several studies[25-37] have focused on using this directed self-assembly technique to synthesize two-dimensional (2D) superstructures of anisotropic particles. These quasi-2D superstructures provide a valuable experimental testbed to study extended electronic properties and have broad technological potential in applications involving thin-film optical and electronic devices[29,30,38,39].

To establish a robust framework to control the self-assembly of nanoparticles at fluid interfaces, a fundamental understanding of the interplay between entropic and enthalpic driving forces is critical. For this purpose, it is instructive to split the study into various complementary steps. The first step would involve simulating a model that isolates entropic effects only. A subsequent step would involve studying a model that also incorporates enthalpic effects like interfacial particle-fluid interactions to assess whether or how specific interactions reinforce or oppose entropic trends. In this work, we focus on the first-step entropic model where nanoparticles are described as perfect, monodisperse, hard polyhedral shapes so that their preferential packing is only dictated by excluded volume interactions. In our model, the particles' centers of mass are pinned to a flat interface such that they cannot translate in a direction perpendicular to that interface, but they are allowed to translate on the surface and rotate freely. The lack of translational freedom in one direction restricts the system of 3D particles to pack in a quasi-2D space and hence we refer to this assembly as 2D confined self-assembly. We use this model to simulate the phase behavior of several shapes belonging to a family of truncated cubes as characterized in ref. 17. Specifically, the shapes studied are cubes, truncated cubes (TCs), truncated cubes with truncation parameter $s=0.4$ (TC4s), cuboctahedrons (COs), truncated octahedrons (TOs) and octahedrons (Octs). These choices are motivated by both their relevance to experimental studies[2,3,29,30] and the availability of reference simulation results for the 3D entropic self-assembly[17].

By way of background, in Section 2 we explain the methodology used to outline the phase behavior of each polyhedron, along with the relevant order parameters used to distinguish different phases. In Section 3, we report the results for the different shapes studied in this work along with the comparisons with the 3D bulk phase behavior and the 2D phase behavior of rounded hard squares[40]. Finally, in Section 4 we provide some concluding remarks.

## 2. Methodology

To confine the self-assembly of the particles in 2D, we fix the $z$-coordinate of the center of mass of each particle and only allow translations in the $x$ and $y$ direction. We perform extensive expansion and compression Monte Carlo (MC) runs in the isothermal-isobaric $NPT$ ensemble ($N$ is number of particles) with standard periodic boundary conditions to map out the equation of state (EoS) for each shape. In our simulations, we assume the particles interact via hard core potential, which forbids particles from overlapping. Each pressure step of expansion/compression involved a run of $3 \times 10^6$ MC cycles in total, with the first $2 \times 10^6$ MC cycles used for equilibration and the latter $10^6$ MC cycles used for production. Each MC cycle consists of an average of $N$ translational, $N$ rotational, $N/10$ flip and 1 area change move attempts. Flip moves attempt to rotate a chosen particle to a random orientation in the plane perpendicular to its present orientation. As indicated before, translational moves are only allowed in $x$ and $y$ direction. At high densities, area moves are allowed to also change the angle between the box axes to relieve internal stresses and avoid the formation of artificial phases due to a fixed square box symmetry. All trial moves are accepted according to the Metropolis criterion[41], which requires ruling out overlaps between any two particles (via the separating axes theorem[42]). The size of the move perturbations is adjusted so as to get acceptance probability values of 0.4, 0.4 and 0.2 for the translation, rotation and volume moves, respectively.

The formation of ordered phases is detected using several global order parameters. The bond order parameters $\psi_4$ and $\psi_6$ are used to detect four-fold and six-fold angular order, respectively. For each particle, identified by $j$, we define a complex number, the local $n$-fold bond orientational order $\varphi_n(\mathbf{r_j})$:

$$\varphi_n(\mathbf{r_j}) = \frac{1}{n_j} \sum_{k=1}^{n_j} \exp(in\theta_{jk}) \qquad (1)$$

for $n = 4$ and 6. In the above equation, $\theta_{jk}$ is the angle made by bond between particle $j$ and its nearest neighbor $k$ with respect to an arbitrary axis, and $n_j$ is the number of nearest neighbors of particle $j$. For $\varphi_6(\mathbf{r_j})$, $n_j$ is calculated using Voronoi tessellation, while for $\varphi_4(\mathbf{r_j})$ the four closest neighbors of each particle are used. The magnitude of $\varphi_n(\mathbf{r_j})$, estimates the value of local $n$-fold bond orientational order, $\varphi_n^j$ for particle $j$. The global bond orientational order, $\psi_n$ for $N$ particles is the magnitude of the average of $\varphi_n(\mathbf{r_j})$ as defined below

$$\psi_n = \left| \frac{1}{N} \sum_{j=1}^{N} \varphi_n(\mathbf{r_j}) \right| \qquad (2)$$

The global particle orientational order in the system is determined via the $P_4$ cubatic order parameter defined as[43]

$$P_4 = \max_{\mathbf{n}} \frac{1}{N} \sum_i P_4(\mathbf{u_i} \cdot \mathbf{n})$$
$$= \max_{\mathbf{n}} \frac{1}{8N} \sum_i (35\cos^4\theta_i(\mathbf{n}) - 30\cos^2\theta_i(\mathbf{n}) + 3) \qquad (3)$$

where, $\mathbf{u_i}$ is the unit vector along a relevant particle axis and $\mathbf{n}$ is a director unit vector which maximizes $P_4$ (see details in ref. 9). The spatial range of orientational order is quantified using the orientational distribution function $I_4(r)$[12], defined as

$$I_4(r) = \frac{1}{8N} \left\langle 35[\mathbf{u}_{ai}(0) \cdot \mathbf{u}_{bi}(r)]^4 - 30[\mathbf{u}_{ai}(0) \cdot \mathbf{u}_{bi}(r)]^2 + 3 \right\rangle \qquad (4)$$

where the average is over all particle pairs and all nine combinations of the axes. Translational order is detected by analyzing the behavior of the radial distribution function $g(r)$, the structure factor, $S(\mathbf{k})$, defined as

$$S(\mathbf{k}) = \frac{1}{N} \left\langle \left[ \sum_{i=1}^{N} \cos(\mathbf{k} \cdot \mathbf{r}_i) \right]^2 + \left[ \sum_{i=1}^{N} \sin(\mathbf{k} \cdot \mathbf{r}_i) \right]^2 \right\rangle, \qquad (5)$$

and the bond orientational correlation function given by
$$g_k(\mathbf{r}) = <\varphi_k(0)\varphi_k(\mathbf{r})> \qquad (6)$$
where $\varphi_k(\mathbf{r})$ is the $k$-fold local bond orientational order at position $\mathbf{r}$.

The regions of phase stability for a given system are outlined by mapping out the EoS via compression and expansion runs. The compression runs were started using a low density isotropic state with negligible translational and orientational order. We continued the compression runs till we obtain the high-density crystalline phase with long ranged translational and orientational order. The resulting phase is then used as a starting point for expansion runs. To detect the points where phase transitions occur, we looked for not only breaks and inflections points in the EoS curves, but also for spots where fluctuations of bond order parameters, defined as

$$\chi_n = N\left(\left\langle \psi_n^2 \right\rangle - \left\langle \psi_n \right\rangle^2 \right) \qquad (7)$$

diverge or have larger peaks[44]. The latter has been found to be less sensitive to finite size effects than the former[40,45]. Since the EoS and transition points found by compression and expansion runs exhibited little hysteresis and we are only interested in outlining approximate boundaries, we did not perform complementary (and costly) free energy calculations.

We use conventional reduced units to report our results for the phase behavior of different shapes. The truncation parameter,[17] $s$ for cubes, TCs, TC4s, COs, TOs, and Octs is 0, (2-√2)/2, 0.4, 1/2, 2/3 and 1 respectively. The reduced pressure, $P^*$ is defined as $P^* = Pa^2/(k_BT)$ where $2a$ is the diagonal of the (imaginary) cube from which the given shape is cut. $k_B$ is Boltzmann constant and $T$ is absolute temperature. The reduced number density is defined as $\eta = Na^2/A$, where $A$ is the total area of the system. We report our results in normalized reduced number density, $\eta^* = \eta/\eta^{crys}$ where $\eta^{crys}$ is the reduced number density of a crystal phase for $N = 1600$ particles at a very high pressure, namely, $P^* = 500$ ($\eta^{crys}$ essentially correspond to the densest packing state).

## 3. Results

### 3.1 Cubes ($s = 0$) and TCs ($s = 0.293$)

For cubes, most simulations entailed a system size of $N = 1600$ particles. The EoS results are shown in Fig.1 including data for a smaller $N = 400$ particle system for comparison. The EoS for the small and large systems from expansion runs matches well for the entire range of densities. We also observed a similar trend for the compression runs of the same system (not shown). At low concentrations, all the order parameters have values close to zero, which is characteristic of the isotropic phase. Around $\eta^* \approx 0.55$, the values of $\psi_4$ and $P_4$ start increasing with pressure while the value of $\psi_6$ stays close to zero. At $\eta^* \approx 0.9$, $\psi_4$ and $P_4$ reach their maximum values, which are consistent with a crystal with square order. A similar isotropic to square phase transition is also observed for a larger system size with $N = 3600$ and has also been observed before in hard squares[46] and rounded hard squares with small corner to length rounding ratios[40]. In the phase behavior of hard squares, ref. 46 shows that the system goes from isotropic to square phase via an intermediate phase referred to as tetratic phase. As described in ref. 40 and ref. 46, the tetratic phase has significant particle orientational and four-fold bond orientational order, but $g(r)$ lacks long-range order and the structure factor $S(\mathbf{k})$ exhibits diffusive peaks with four-fold symmetry. These tetratic-like properties are observed in our case for a region intermediate between the isotropic and square phases. As shown in Fig. 1. for $0.6 < \eta^* < 0.7$, one observes that $0.4 < \psi_4 < 0.8$, and $0.4 < P_4 < 0.45$ whiled the coupling between particle orientations and (nearest neighbor) bond orientations also increases rapidly as detected by the cross parameter $\psi_4 P_4$,. As shown in Fig. 2, at these conditions $S(\mathbf{k})$ exhibits diffuse peaks with four fold symmetry and $g(r)$ shows a lack of long-range structure consistent with tetratic-like behavior. For comparison, we also show in Fig. 2 plots of the structure factor plot and $g(r)$ for the square phase. The lack of any translational degrees of freedom in $z$-direction in our model renders the behavior of hard cubes similar to that of hard squares. We also compare this self-assembly of cubes confined in 2D to the self-assembly of cubes in the bulk in 3D[12,17]. Besides the expected differences in the symmetry of the densest packings ( square in 2D and cubic in 3D), we also find that at least for the system sizes studied here cubes undergo a continuous phase transition from isotropic to square phase which differs from the first order phase transition from isotropic to simple cubic phase observed in bulk 3D self-assembly. The pseudo 2D phase behavior of TCs is very similar to that of cubes and is described in the supplementary information[47].

### 3.2 TC4s ($s = 0.4$)

As shown in Fig. 3, a system of $N = 1600$ T4Cs transitions from isotropic to square phase via two intermediate mesophases. Firstly, the system undergoes a transition from isotropic to a phase with partial six fold bond orientational order and no particle orientational order, which, on further increasing the pressure, loses its six fold symmetry and gains both four fold bond orientational order and particle orientational order. Further compression of the system results in a crystalline square phase. We observe a similar phase behavior for a larger system sizes with $N = 3600$ (results not shown). To better understand the two intermediate phases, we first look at the phase with $\psi_6 \approx 0.45$ observed for $0.65 < \eta^* < 0.69$ and find that its $S(\mathbf{k})$ shows diffuse peaks with six-fold symmetry (see Fig. 4b). The corresponding correlation function $g_6(r)$ (Fig. 5) shows long-range partial hexatic order whereas $g(r)$ shows a quick decay of the peak amplitudes with distance, indicative of short-range translational order. These properties are consistent with those of the hexatic phase observed in the case of hard discs. We note, however, that hexatic character could only be unambiguously determined with a scaling analysis involving much larger system sizes than those used here, a task that lies beyond the scope of this work. For the other intermediate phase observed in $0.71 < \eta^* < 0.76$, $S(\mathbf{k})$ exhibits diffuse four-fold peaks, $g_6(r)$ and $I_4(r)$ reveal long-range partial tetratic bond orientational and particle orientational order, and $g(r)$ indicates a short-range translational order (see Figs. 4 and 5). This intermediate phase has similar properties to those of the transitional phase observed before in cubes and TCs when the system transitions from isotropic to square phase. The absence of breaks in order parameters and $\eta^*$ vs. pressure as the system transitions between phases suggests that the transitions are continuous. Interestingly, similar to the phases observed here for TC4s confined in 2D, the bulk self-assembly of TC4s in 3D also exhibits an intermediate mesophase with six-fold symmetry and a densest packing phase with a different symmetry.[17]

### 3.3 COs ($s = 0.5$)

The EoS for $N = 1600$ COs is shown in Fig. 6. The system undergoes a phase transition from isotropic to a hexagonal rotator phase characterized by high six-fold bond orientational order and negligible particle orientational order. This transition is accompanied by discontinuities in the values of $\eta^*$ and $\psi_6$, suggesting that the transition would be first order. The hexagonal rotator phase in the range of $0.67 < \eta^* < 0.77$ shows sharp $S(\mathbf{k})$ peaks of hexagonal symmetry which is in line with the long-range ordering behavior detected in $g(r)$ (see Fig. 7). By further increasing the density, the system transitions from a hexagonal rotator phase to a crystal phase having high particle orientational order and a distorted translational symmetry. The $g(r)$ of the crystal phase has a solid like behavior (peaks persisting to large distances) as shown in Fig. 7 but the $S(\mathbf{k})$ plot shows that system has neither fully six-fold nor four-fold symmetry. To better understand this crystal phase, we calculate $\varphi_4^j$ and $\varphi_6^j$ for each particle and show in Fig. 10 a snapshot of the system at $P^*=59.4$ by coloring the particles based on their (local) values of $\varphi_4^j$ and $\varphi_6^j$ [as per Eq. (1)]. The system has mainly hexagonal-like character which is consistent with a high global $\psi_6 \sim 0.85$, but also has layers of four-fold symmetry, indicating some localized interlocking between phases with six fold and four fold symmetry. A similar interlocking between phases has also been observed in the simulated crystal phase of COs assembling in 3D[17]. Also, COs in 3D exhibit isotropic, hexagonal rotator phase and a crystal phase with a distorted lattice upon compression, a sequence of phases that is loosely "equivalent" to that we observed in 2D.

### 3.4 TOs ($s = 2/3$)

The phase behavior of TOs for a system of $N = 1600$ particles is shown in Fig. 9. As can be observed, the EoS for $N=400$ and $N=1600$ agree very well and reveal two phase transitions. The first one, observed at $\eta^* \approx 0.6$, is the transformation from an isotropic phase to hexagonal rotator phase characterized by a large increase in $\psi_6$. The second transition observed at $\eta^* \approx 0.83$ is the transformation from hexagonal rotator phase to rhombic phase, that latter characterized by large values of $\psi_4$, $\psi_6$ and $P_4$. Figure 10 shows representative snapshots of both hexagonal rotator phase and rhombic phase including the corresponding structure factors. The difference in hexagonal rotator phase and rhombic phase can also be observed by plotting the distributions of angles, $\alpha$, made by the vector joining an $i^{th}$ particle and its closest neighbor with the vector joining an $i^{th}$ particle and its next five closest neighbors (where neighbors are identified via Voronoi tessellation). As expected, for the hexagonal rotator phase this angular distribution peaks at 60, 120 and 180 degrees. In contrast, for the rhombic phase we see 5 peaks at around 55, 70, 110, 125 and 180 degrees. The discontinuities in order parameters and $\eta^*$ for the two phase transitions suggest that they are both first order.

The phases observed for this case are similar to the 2D phases observed for rounded hard squares with high corner to length rounding ratios[40]. Similar to the confined case studied here, a transition akin to isotropic-to-rotator phase is also observed in the self-assembly of TOs in 3D[12,17], but unlike our observation of particles abruptly aligning to transform from rotator phase to rhombic phase, a continuous transition from orientationally disordered rotator phase to bcc phase is observed in 3D.

### 3.5 Octs ($s = 1$)

The phase behavior of Octs for $N = 1600$ particles is shown in Fig. 11. The discontinuity in the order parameter values of $\psi_4$, $\psi_6$ and $P_4$ and $\eta^*$ around $\eta^* \approx 0.67$ suggests that the system transforms into a hexagonal crystal phase via a first-order transition. A representative snapshot of the hexagonal crystal phase along with its corresponding structure factor and radial distribution function are shown in Fig. 12. We observe sharp $S(\mathbf{k})$ peaks with six-fold symmetry for the hexagonal crystal phase, consistent with the long range structural order detected in $g(r)$. Over a short density range before $\eta^* \approx 0.67$ the system has no particle orientational order (low $P_4$ values) but has non negligible values of $\psi_4$ and $\psi_6$, which we associate with a novel mesophase that we'll denote as a "dimorphic" rotator. The same phase is observed in expansion and compression runs for $N = 3600$ particles. As shown in Fig.12 for $N = 3600$ particles, the structure factor plot of this mesophase reveals diffuse peaks with no clear symmetry while $g(r)$ shows small but persistent peak at long distances. To try to further characterize the structure of this phase, we visually analyzed sample configurations by coloring the particles based on their individual values of $\varphi_4^j$ and $\varphi_6^j$ [see Eq. (1)]. As Figure 13 shows, this mesophase has a grainy character where two kinds of "grains" intermingle (hence the name dimorphic): some are patches of particles having high six-fold symmetry and low four-fold symmetry, while others are patches of particles having high four-fold and low six-fold symmetry. While this dimorphic phase bears some similarity to the so-called polycrystalline phase detected for a small range of concentrations and roundedness parameter in rounded squares[40], it seems to have less translational order.

### 3.6 General trends of behavior

Figure 14 summarizes the phase behavior for all the shapes studied here. As the value of $s$ is increased, the range of densities at which a phase with six-fold symmetry is stable increases and the reverse trend ensues for a phase with four-fold symmetry. Based on this approximate trend, one would expect that the two-body potential of mean force (*PMF*) of the different shapes may reveal some signs for a stronger disc-like behavior with increasing $s$. Hence, we estimate one-dimensional and two-dimensional *PMFs* at different values of $s$ using the procedure mentioned in ref. 47. As expected, the one-dimensional *PMF(r)* shape (Fig. S3 in the supplementary material[47]) has a more compact repulsive tail and shorter range (i.e., more disc-like character) for shapes with larger $s$, although this trend is somewhat reversed between COs and Octs. The *PMF(r)* shape of Octs for larger values of $r$ ($>1.15$) is intermediate between those for COs and TCs, which would be consistent with an intermediate proclivity toward disk-like and square-like behavior as embodied by the novel dimorphic phase formed by Octs. The two-dimensional $PMF(x,y)$ (Fig. S4 in the supplementary material[47]) allows visualization of the asymmetry in the effective two-particle interactions along their relative (in-plane) orientations. For any such orientation, the value of $\beta PMF(x,y)$ decays faster over distance from large repulsion ($\beta PMF(x,y) \geq 5$) to no repulsion for shapes with larger $s$, which is consistent with the trends seen in $PMF(r)$. Also, the TOs' $PMF(x,y)$ has a marked circular-like symmetry which is consistent with the hexagonal rotator phase that was detected for a wide range of densities.

Besides summarizing the wide variety of phases we observe for different values of $s$, Figure 14 also gives a sense for how such phases are related across shapes and provides some clues as to how different phase types could be interpolated (for shapes whose $s$ we did not simulate). We observe that for each phase observed in our quasi 2D systems, a counter part is almost always identifiable in the 3D bulk phase behavior from ref. 17, in particular, the number and sequence of mesophases and solid phases match well for the systems simulated. Of course that lattice symmetries in the mesophases and crystal phases are necessarily different between the 3D and quasi-2D systems; further, the character of the phase transitions (first-order or continuous) either did not always corresponds or could not be unambiguously determined in our systems. Table 1 shows the approximate phase correspondences (ignoring the isotropic phase). The only mesophase that appears to have no counterpart in the 3D case is the dimorphic mesophase we found in Octs.

The phase behavior of our systems also has some similarities with that of rounded squares. Arguably, the degree of truncation embodied by each shape could potentially be mapped to an effective degree of square roundedness, corresponding perhaps to the average projection of the shape onto the 2D pinning plane or to some key similarity in their $PMF$s. However, such a mapping would be flawed, since such a roundedness parameter would depend on concentration (which affect the azimuthal rotational states accessible to a particle) for a given particle shape. Nonetheless we observe some correlation between truncation parameter, $s$ and corner-rounding-to-length ratio, $\zeta$ (a parameter used to characterize the roundness of particles in ref. 40). For larger values of $s$ and $\zeta$, both types of systems show more disc-like behavior whereas at smaller values of $s$ and $\zeta$, the systems exhibit square-like phase behavior. Along with this general trend, we also observe a marked similarity in the phase behavior for some values of $s$, like for the $s = 0$ and $\zeta = 0$ cases, both systems go from isotropic to a tetratic phase and then to a square phase. Also, when both $s$ and $\zeta$ have a value of 2/3, the systems undergo a transition from isotropic to hexagonal rotator phase followed by a transition to a rhombic phase.

## 4. Conclusions and Discussion

In this work, we have simulated the confined 2D phase behavior of 6 different shapes, from cubes to octahedra, which belong to a family of truncated cubes often encountered in nanoparticle synthesis. The phase behavior of different shapes shows diversity in the type of mesophases and crystalline phases that form as concentration increases. Shapes with small values of $s$ (cubes and TCs) shows a square-like behavior, where upon compression a tetratic-like phase is first observed, followed by a square phase. On the other hand, shapes with large values of $s$ (COs, TOs and Octs) show phases with a disc-like character, where upon compression form a rotator phase followed by a crystal phase with both rotator and crystal phases exhibiting an increasing (decreasing) extent of hexagonal (square) symmetry as $s$ increases. Interestingly, for Octs a mesophase with patches of six-fold and four-fold symmetry is observed for a small range of densities before it transitions to a hexagonal crystal. For the intermediate value of $s = 0.4$ (TC4s), we observe both square-like and disc-like packing in the phase behavior. Indeed, TC4s undergo a transition from isotropic to hexatic-like phase followed by a tetratic-like phase, which finally transforms into a square phase. While breaks or inflexion points in order parameters and number density along an equation of state are suggestive of the potential character the phase transitions involved, further finite-size scaling studies will be required to clarify the exact nature of these transitions.

The present work constitutes a first step towards the goal of fundamentally understanding the self-assembly of polyhedral nanoparticles at fluid-fluid interfaces. The phase behavior obtained here for purely entropic self-assembly provides a baseline for later comparisons; e.g., to understand additional entropic effects like those related to size and shape polydispersity (ubiquitous in experimental systems) and to facilitate the partial decoupling of effects brought about by the addition of enthalpic interactions. Despite the simplicity of the entropic model, some of the phases and mesophases observed here have been observed in various experiments[29-38] (and in our ongoing work), and while a direct comparison is still not possible, this concurrence nonetheless illustrates that mesophasic behavior is ubiquitous in experiments and it can occur across different types of particle interactions, potentially foretelling the preponderance of steric effects on structure. Furthermore, the entropic phase behavior found here may also prove to provide best-case scenarios for ordered assembly, which future experiments may aim to approach.

One of the important enthalpic contributions in the directed self-assembly of particles at the interface between fluids 1 and 2 is the interfacial energy originating from the interactions between particle and fluid 1, particle and fluid 2, and between the two fluids. To build a model with such additional enthalpic interfacial energy, one first needs to understand the equilibrium adsorption configuration of a single particle at the fluid 1/fluid 2 interface based on the properties of the two fluids and a particle. As suggested in refs. 30 and 31, this can be simplified by using a Pieranski potential[24] where a 3D interfacial energy landscape can be obtained based on the polar angles, azimuthal angles and immersion depths of a particle relative to the fluid-fluid interface. The authors of ref. 31 studied the interfacial self-assembly of hexagonal bipyramid and bifrustum shaped particles into 2D superstructures by performing MC simulations where the immersion depth and the polar angle of a particle were set according to the values estimated for the equilibrium adsorption configurations. Such constraints could be relaxed in future studies to more faithfully describe the role of metastable polar-angle particle orientations. Furthermore, enthalpic particle-particle interactions can be incorporated to reflect ligand-mediated interactions present in some experimental systems, which could crucially affect the assembly behavior. Work along these lines is already under way.

**Acknowledgements**.
The authors acknowledge funding support from a SEED Grant provided by NSF MRSEC award to Cornell, Grant DMR-1120296. The authors are also grateful to Mihir Khadilkar and Kevin Whitham for useful exchanges.

**TABLES**

Table 1. Summary of phase behavior and comparison to 3D bulk systems. sc = simple cubic, dsc = distorted sc, MI = monointerlocking, bct = body-centered tetragonal, dbct = distorted bct.

| Shape | Quasi-2D case | 3D bulk case [11] |
|---|---|---|
| Cubes | Tetratic-like square | Cubic mesophase* sc |
| Truncated cubes (TCs) | Tetratic-like Square | sc dsc (C1) |
| TC4s | Hexatic-like rotator Tetratic-like Square | Plastic hcp dsc (C0) MI dsc |
| Cuboctahedra (COs) | Hexagonal rotator Distorted hexagonal | Plastic bct dbct0 |
| Truncated Octahedra (TOs) | Hexagonal rotator Rhombic | Plastic bct** dbct1 |
| Octahedra (Octs) | Dimorphic rotator Hexagonal crystal | - Minkowski crystal |

*) For a narrow range of densities near the ordering transition, the particles in the cubic phase have liquid-like mobilities which could be associated with mesophasic behavior[12,13].

**) Rotator phase behavior for TOs has been detected in refs. 12 and 21 although it continuously transitions into a crystalline phase as density increases.

**FIGURES**

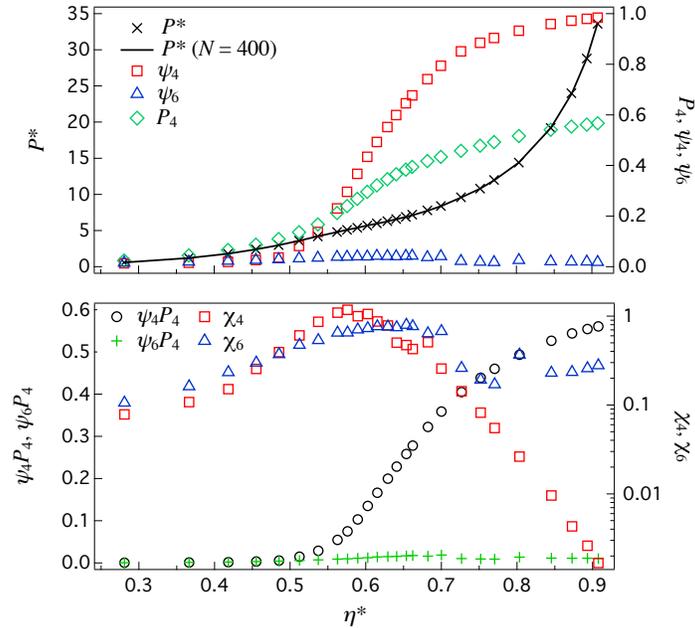

**Fig. 1** Equation of state for 1600 cubes obtained by expansion runs. (Top Panel) The pressure $P^*$, bond orientational order parameters $\psi_4$ and $\psi_6$, and orientational order parameter $P_4$ as a function of reduced number density $\eta^*$. For comparison, the equation of state curve obtained from the system of $N = 400$ particles is also shown. (Bottom Panel) Cross order parameters $\psi_4 P_4$ and $\psi_6 P_4$ and susceptibilities of the bond order parameters $\chi_4$ and $\chi_6$ as a function of $\eta^*$. The value of $\eta^{crys}$ is estimated as 0.745.

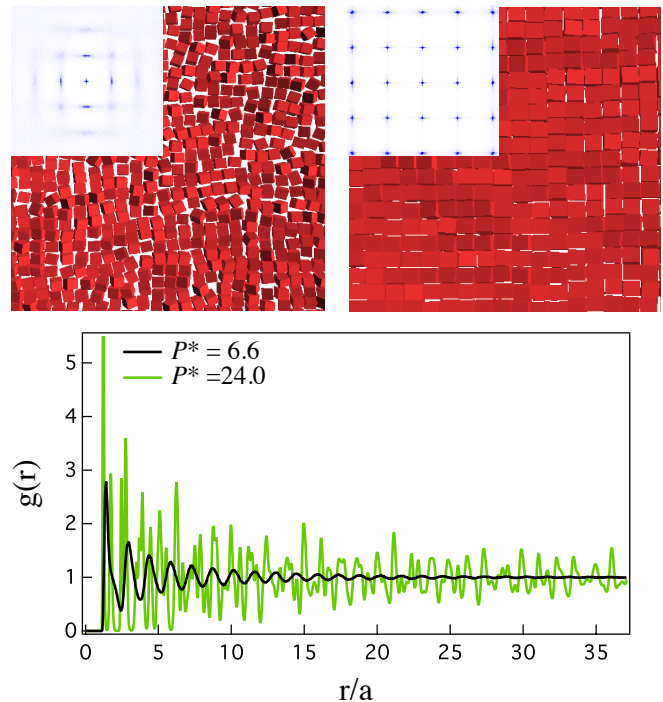

**Fig. 2** (Top Panel) Representative snapshots and corresponding structure factors for a system of $N=3600$ cubes at (left) $P^* = 6.6$ (tetratic-like phase) and (right) $P^* = 24.0$ (square phase). (Bottom Panel) The radial distribution function for a system of $N = 3600$ cubes at the same two pressures.

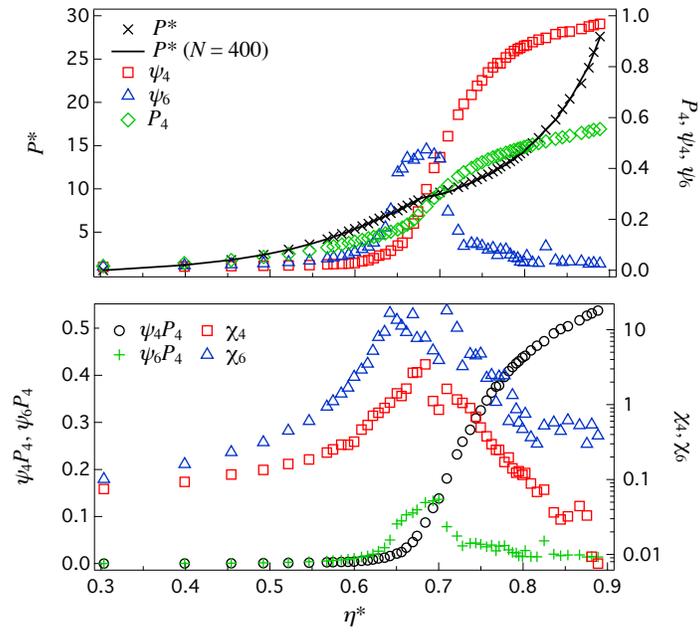

**Fig. 3** Equation of state for 1600 TC4s obtained by expansion runs. Legend as in Fig. 1. The value of $\eta^{crys}$ is estimated as 0.745.

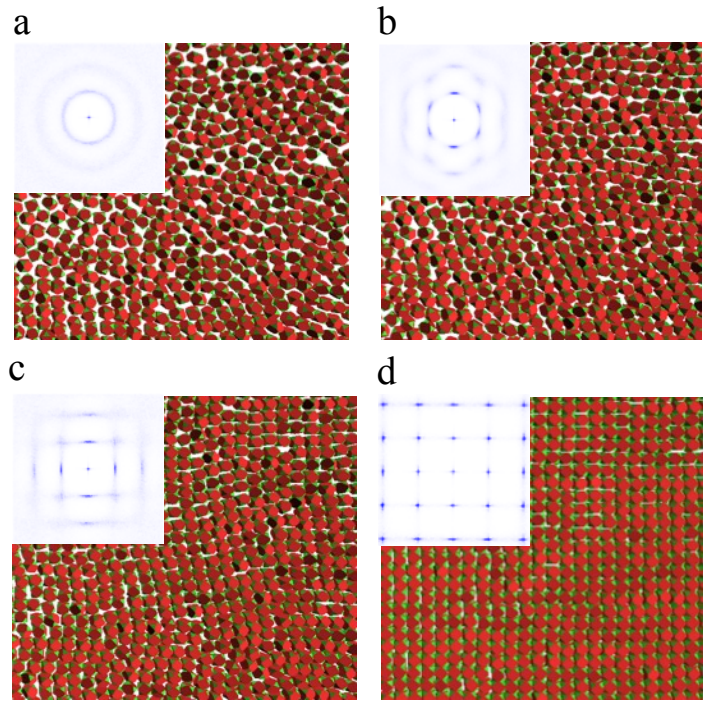

**Fig. 4** Snapshots and corresponding structure factors for $N = 3600$ TC4s at (a) $P^* = 6.0$ (isotropic phase), (b) $P^* = 8.4$ (hexatic-like phase), (c) $P^* = 10.8$ (tetratic-like phase) and (d) $P^* = 24.0$ (square phase).

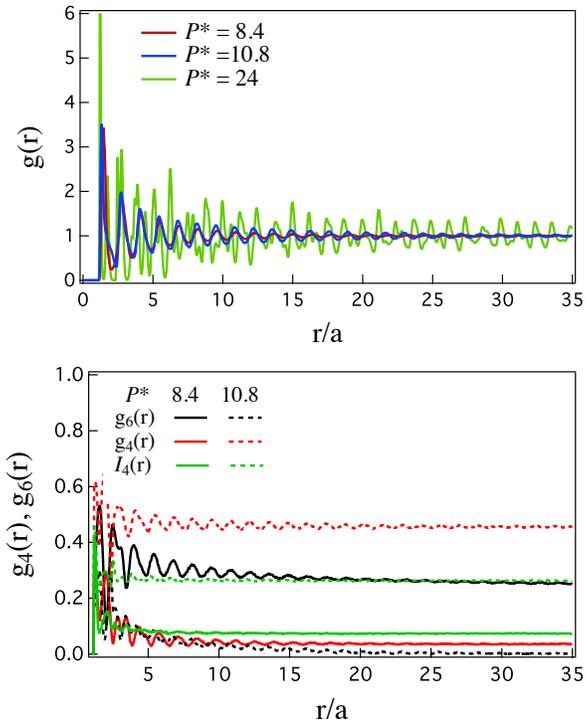

**Fig. 5** Correlation functions for the system of $N = 3600$ TC4s. (Top Panel) The radial distribution function g(r) at $P^* = 8.4$ (hexatic-like phase), 10.8 (tetratic-like phase) and 24 (square phase) are shown. (Bottom Panel) The bond order correlation functions, $g_6(r)$ and $g_4(r)$, and orientational correlation function, $I_4(r)$ at $P^* = 8.4$ and 10.8 are shown.

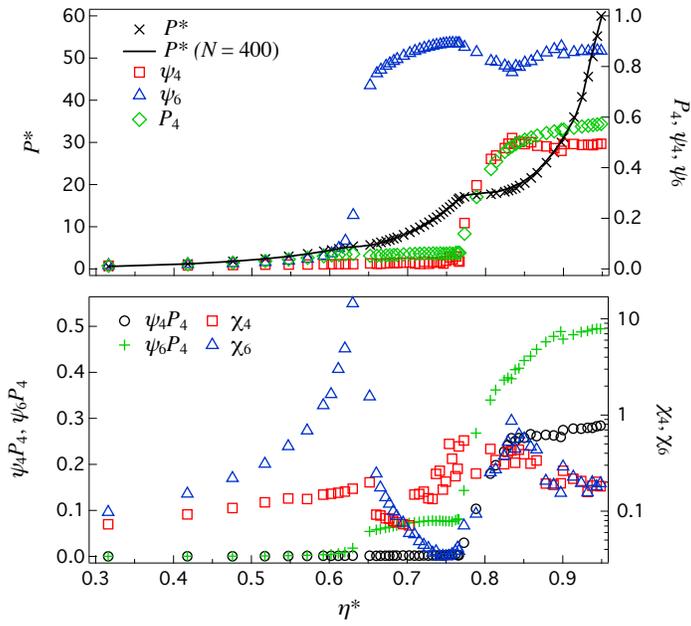

**Fig. 6** Equation of state for 1600 COs obtained by expansion runs. Legend as in Fig. 1. The value of $\eta^{crys}$ is estimated as 0.745.

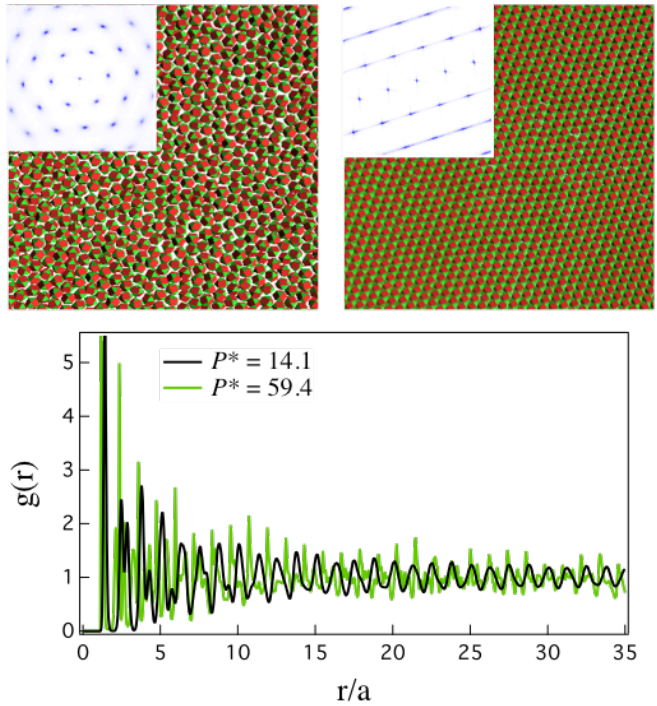

**Fig. 7** (Top Panel) Representative snapshots and corresponding structure factors for a system of $N=3600$ COs at (left) $P^*= 14.1$ (hexagonal rotator phase) and (right) $P^* = 59.4$ (distorted crystal phase). (Bottom Panel) The radial distribution function for a system of $N = 3600$ COs at the same two pressures.

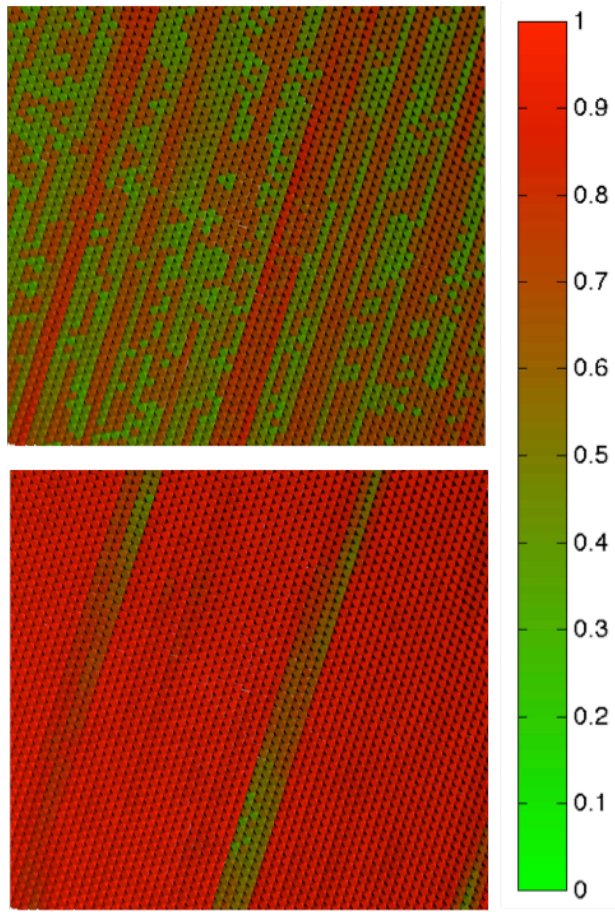

**Fig. 8** Snapshot of the distorted crystal phase for the system of $N = 3600$ COs at $P^*=59.4$. (Top Panel) Particles are colored with respect to the local value of $\varphi_4^j$. (Bottom Panel) Particles are colored with respect to the local value of $\varphi_6^j$.

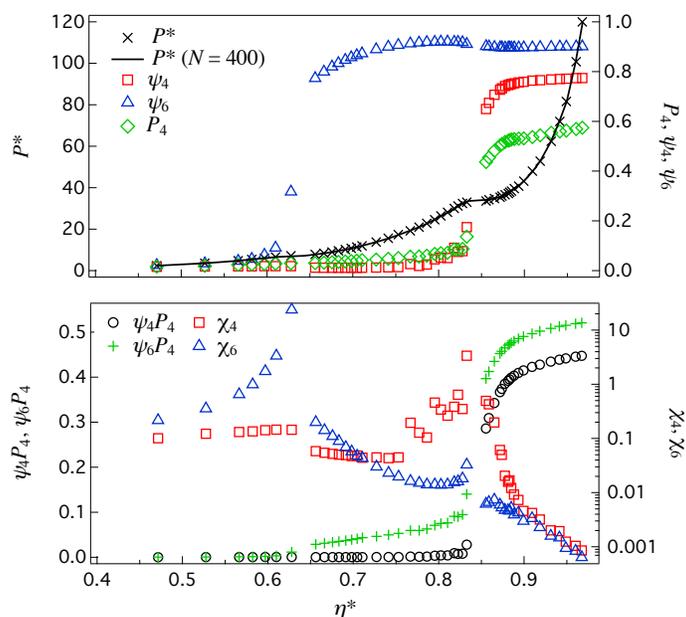

**Fig. 9** Equation of state for 1600 TOs obtained by expansion runs. Legend as in Fig. 1. The value of $\eta^{crys}$ is estimated as 1.05.

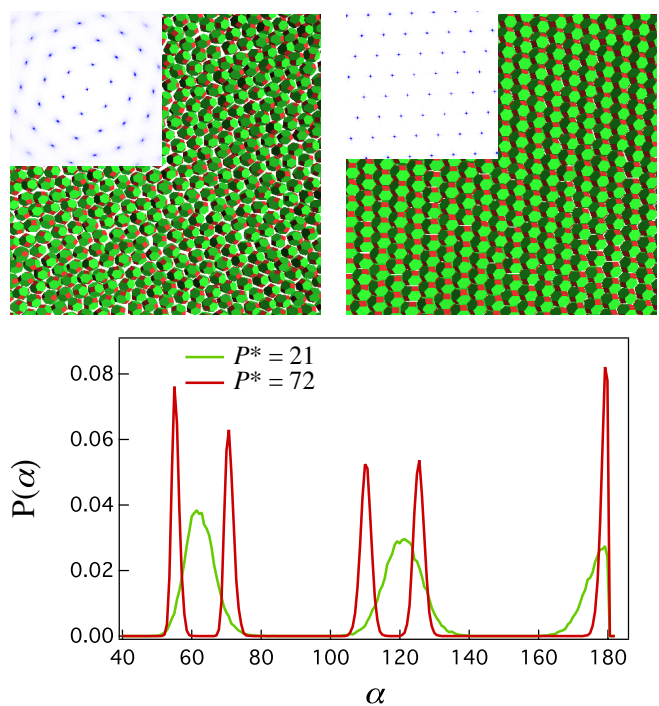

**Fig. 10** (Top Panel) Representative snapshots and corresponding structure factors for a system of $N=1600$ TOs at (left) $P^* = 21$ (hexagonal rotator phase) and (right) $P^* = 72$ (rhombic phase). (Bottom Panel) Probability density function of angle, $\alpha$ for a system of $N = 1600$ TOs at the same two pressures.

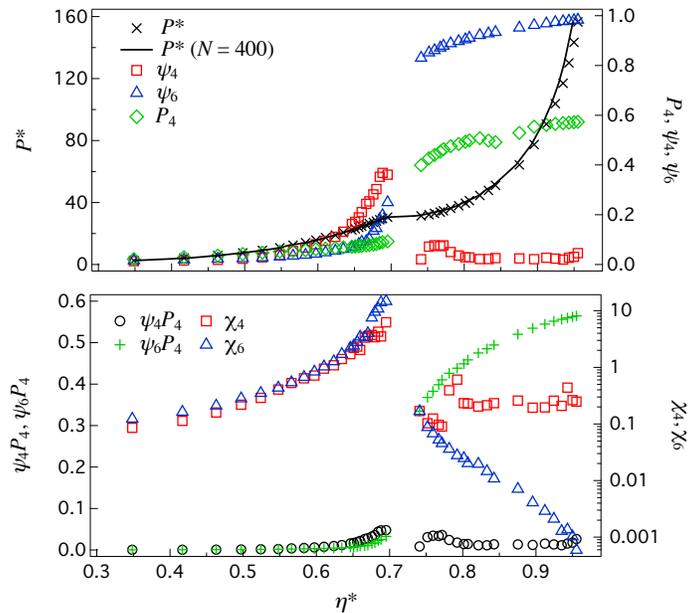

**Fig. 11** Equation of state for 1600 Octs obtained by expansion runs. Legend as in Fig. 1. The value of $\eta^{crys}$ is estimated as 2.247.

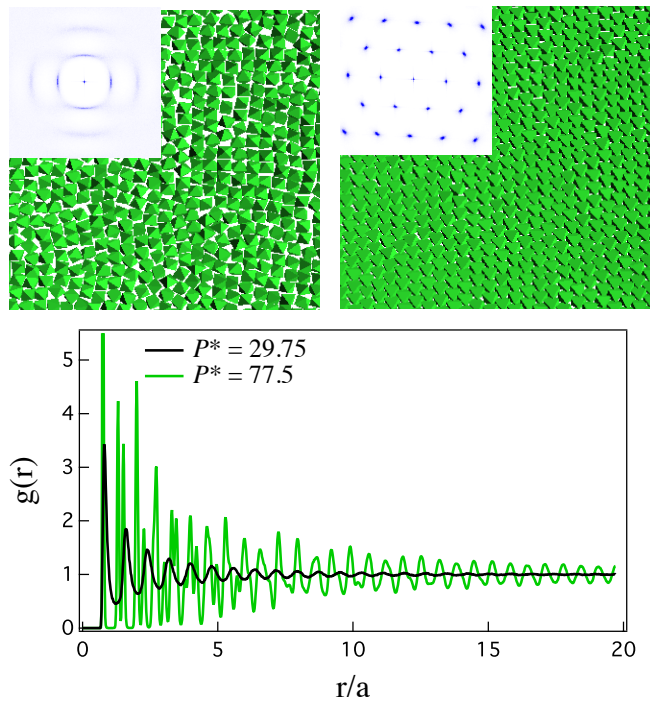

**Fig. 12** (Top Panel) Representative snapshots and corresponding structure factors for a system of $N=3600$ Octs at (left) $P^*= 29.75$ (mixed phase) and (right) $P^* = 77.5$ (hexagonal crystal phase). (Bottom Panel) The radial distribution function for a system of $N = 3600$ Octs at the same two pressures.

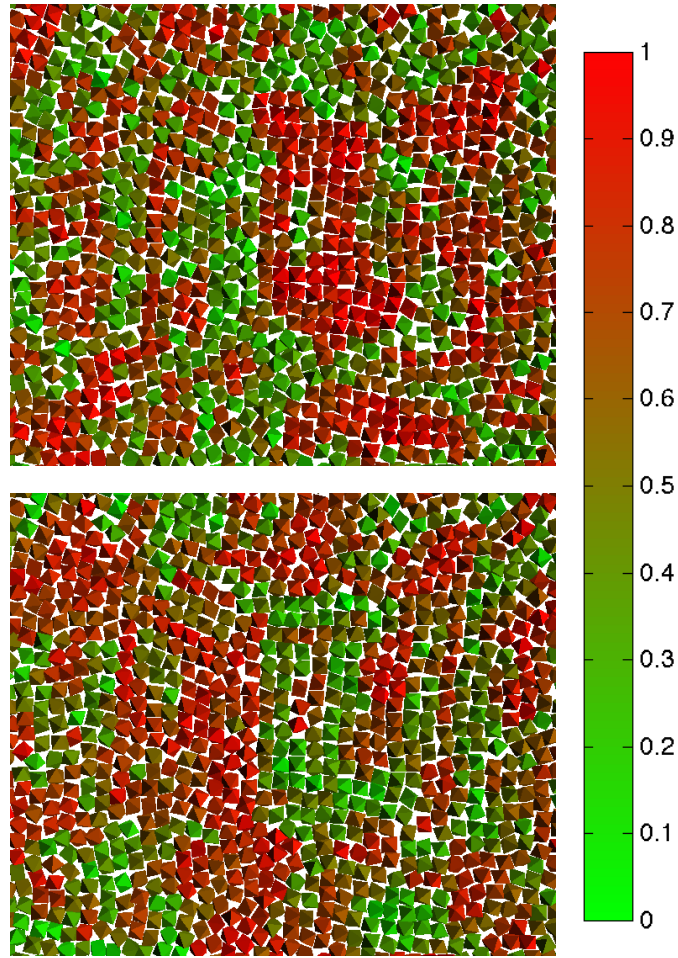

**Fig. 13** Snapshot of the phase observed at $P^*= 29.75$ for the system of $N = 3600$ Octs. (Top Panel) Particles are colored with respect to the local value of $\varphi_4^j$. (Bottom Panel) Particles are colored with respect to the local value of $\varphi_6^j$.

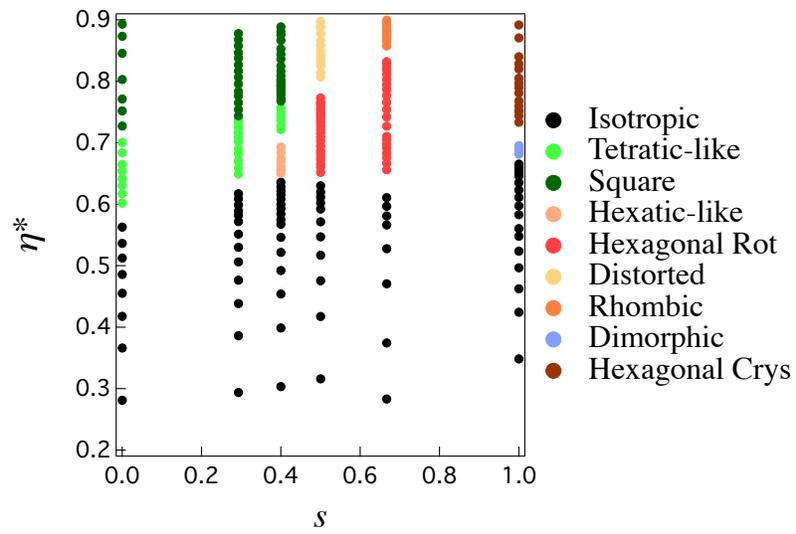

**Fig. 14** Concentration vs. truncation parameter phase diagram for different shapes obtained from MC expansion runs for systems with $N = 1600$ particles.